\documentclass[a4paper,11pt]{article}
\usepackage{pos}
\usepackage{here}
\usepackage{xspace}
\usepackage{multirow}
\usepackage{booktabs}
\usepackage{slashed}

\usepackage{colortbl}
\usepackage{xcolor}

\newcommand{\GeV}{{\ensuremath\,\rm GeV}\xspace}
\newcommand{\TeV}{{\ensuremath\,\rm TeV}\xspace}
\newcommand{\pb}{{\ensuremath\,\rm pb}\xspace}

\title{Search for Light Scalars in the TRSM at the LHC}

\author[a]{Aman Desai}
\author[b]{Kristin Lohwasser}
\author*[c]{Mohamed Ouchemhou}
\author[c]{Tania Robens}
\author[d,e]{Prasenjit Sanyal}

\affiliation[a]{Department of Physics, Adelaide University, North Terrace, Adelaide, SA 5005, Australia}

\affiliation[b]{School of Mathematical and Physical Sciences, University of Sheffield, Hicks Building, Hounsfield Road, Sheffield, S3
7RH, UK}

\affiliation[c]{Ruder Boskovic Institute, Bijenicka cesta 54, 10000 Zagreb, Croatia}

\affiliation[d]{Center for Quantum Spacetime, Sogang University, Seoul 121-742, South Korea}

\affiliation[e]{Department of Physics, Sogang University, Seoul 121-742, South Korea}

\emailAdd{aman.desai@adelaide.edu.au}
\emailAdd{kristin.lohwasser@cern.ch}
\emailAdd{Mohamed.Ouchemhou@irb.hr}
\emailAdd{trobens@irb.hr}
\emailAdd{prasenjit.sanyal01@gmail.com}

\abstract{

We study the production of Beyond the Standard Model light scalar states in association with a vector boson  ($Vh_2$, with $V = W^\pm, Z$) at the LHC. We consider the scenario where the Standard Model scalar sector is extended by two real scalar singlets, where these additional scalars have mass $ M_i \leq M_{h_{125}}$. In this work, the scalar boson $h_2$ decays via $h_2 \to h_1 h_1 \to 4b$, while the associated vector boson decays either into a pair of oppositely charged leptons or into a single charged lepton and a neutrino. We analyze the signal using LHC detector parameterizations and evaluate its statistical significance at a center-of-mass energy of 13.6~TeV for integrated luminosities of 300~fb$^{-1}$ and 3000~fb$^{-1}$ corresponding to the LHC Run 3 and High Luminosity LHC, respectively. Our preliminary results indicate promising discovery prospects for this channel serving as a complementary probe of extended scalar sectors.\\
RBI-ThPhys-2026-04, COMETA-2026-04
}

\FullConference{Proceedings of the Corfu Summer Institute 2025 "School and Workshops on Elementary Particle Physics and Gravity" (CORFU2025)\\
24 August - 03 September, 2025\\
Corfu, Greece\\}

\begin{document}
\maketitle

\section{Introduction}

In this proceeding contribution, we investigate a new physics scenario that provides several new scalar states, and try to estimate the discovery potential for this model at hadron colliders. Models with scalar extensions are of interest for various reasons. A dogmatic approach could be to say that there is a priori no reason that there is only one scalar boson realized in nature. Furthermore, models with extended scalar sectors can solve several puzzles of the Standard Model (SM), as e.g. the existence of dark energy, the evolution of the universe and CP violation, as well as the metastability of the SM potential. We refer the reader to \cite{Robens:2025nev} for a recent review.

A wide class of BSM models, motivated by the shortcomings of the SM, predicts additional scalar or pseudoscalar states. A priori, there is no limit to extend the scalar sector of the SM as long as all current theoretical and experimental bounds are fulfilled. 
In this work, we focus on an extension of the SM with two real scalar singlets (TRSM)~\cite{Robens:2019kga,Robens:2022nnw}, which gives rise to three CP-even Higgs states, $h_1$, $h_2$, and $h_3$. In \cite{Robens:2019kga}, several benchmark planes were proposed that were by that time unexplored by the experimental collaborations. We here focus on a scenario where the SM-like scalar is the heaviest of the three CP-even neutral scalar states. In this benchmark plane, we explore the associated production of a light CP-even scalar with an electroweak gauge boson, $V\,h_2$, followed by the decay
$h_2 \to h_1 h_1 \to b\bar{b}b\bar{b}$. 
This leads to final states with four $b$-jets in association with either a charged lepton and missing energy or a pair of opposite-sign leptons. Other related experimental searches, mostly with the SM like scalar as a mediator, can be found in Refs.~\cite{ATLAS:2020ahi, ATLAS:2024itc,CMS:2022xxa, CMS:2024zfv}.

Current searches on the exotic decay $h \to aa$, where $a$ is a light scalar or pseudoscalar, have been extensively performed by the ATLAS and CMS collaborations. These include fully hadronic final states such as $4b$~\cite{ATLAS:2025rfm,ATLAS:2018pvw,CMS:2024zfv}, as well as mixed channels like $\mu\mu bb$ and $\tau\tau bb$~\cite{CMS:2024uru}. Complementary searches target leptonic and photonic signatures, including $bb\tau\tau$~\cite{ATLAS:2024vpj}, $\gamma\gamma\tau\tau$~\cite{ATLAS:2024nnm}, and the clean $4\gamma$ final state~\cite{ATLAS:2023ian}, which is particularly sensitive to very light pseudoscalars. Together, these channels provide broad coverage of the parameter space for $m_a \lesssim m_h/2$.

We first explore the parameter space of the TRSM by imposing the relevant theoretical constraints and experimental bounds.
Based on this analysis, we define three representative benchmark points, which are subsequently studied at the LHC with a center-of-mass energy of $13.6$~TeV and an integrated luminosity of 300~fb$^{-1}$, as well as in projections for the High-Luminosity LHC at the same center-of-mass energy.
Finally, we present the discovery prospects of this channel, showing that it provides a complementary and sensitive probe of extended scalar sectors.

\section{Extended Scalar Sector with Two Singlets}

The Two Real Singlet Model ~\cite{Robens:2019kga,Robens:2022nnw} (see also \cite{Robens:2025tew} for the most recent updates) represents one of the simplest extensions of the Standard Model (SM) scalar sector by introducing two additional real singlet fields $S$ and $X$, in addition to the SM Higgs doublet $\Phi$. To reduce the number of free parameters and ensure the stability of the scalar potential, the framework is endowed with two discrete $\mathbb{Z}_2$ symmetries, acting on the fields as
\begin{align}
	\mathbb{Z}_2^S &: S \to -S, \quad X \to X, \\
	\mathbb{Z}_2^X &: S \to S, \quad X \to -X.
\end{align}
All SM fields transform evenly under the above novel symmetries.

Under the symmetry group $\mathbb{Z}_2^S \otimes \mathbb{Z}_2^X$, the most general renormalizable and invariant scalar potential can be written as
\begin{align}
	V(\Phi,S,X) =\;&
	-\mu_\Phi^2\, \Phi^\dagger \Phi
	+ \lambda_\Phi (\Phi^\dagger \Phi)^2
	+ \frac{1}{2}\mu_{S}^2 S^2
	+ \frac{1}{2}\mu_X^2 X^2 
	+ \frac{1}{4}\lambda_{S} S^4
	+ \frac{1}{4}\lambda_X X^4 \nonumber \\
	&+ \frac{1}{2}\lambda_{S X} S^2 X^2
	+ \frac{1}{2}\lambda_{\Phi S} (\Phi^\dagger \Phi) S^2
	+ \frac{1}{2}\lambda_{\Phi X} (\Phi^\dagger \Phi) X^2 \, .
	\label{potential}
\end{align}

The scalar potential in Eq.~\eqref{potential} contains nine real parameters, corresponding to mass terms and quartic couplings, which can later be expressed in terms of physical observables. After electroweak symmetry breaking, the scalar fields may acquire vacuum expectation values (VEVs). 
Expanding the fields around their VEVs, we write
\begin{equation}
	\Phi =
	\begin{pmatrix}
		0 \\
		\dfrac{v + \phi_h}{\sqrt{2}}
	\end{pmatrix}, \qquad
	S = \frac{v_S + \phi_S}{\sqrt{2}}, \qquad
	X = \frac{v_X + \phi_X}{\sqrt{2}} \, .
\end{equation}

In this work, we focus on the scenario where both singlets acquire non-vanishing VEVs. 
This leads to mixing among the doublet and singlet fields and results in a rich and phenomenologically interesting scalar spectrum. The physical CP-even mass eigenstates $h_i$ $(i=1,2,3)$ are related to the gauge eigenstates through an orthogonal $3\times3$ rotation matrix $\mathcal{R}$:
\begin{equation}
	(h_1, h_2, h_3)^T = \mathcal{R}\, (\phi_h, \phi_S, \phi_X)^T .
\end{equation}
The rotation matrix can be parametrized in terms of three mixing angles $\theta_{ij}$ $(ij = hS,\, hX,\, SX)$. In this work we use a mass hierarchy implying $M_1\,\leq\,M_2\,\leq\,M_3$.

The gauge singlets do not contribute to electroweak symmetry breaking, which proceeds as in the SM. The Higgs doublet VEV is fixed to its SM value, $v = v_{\rm SM} \sim 246~\mathrm{GeV}$. 
One of the mass eigenstates is identified with the observed Higgs boson with mass around $125$~GeV. 
In our analysis, we choose $h_3 \equiv h_{\rm SM}$. The model can then be parametrized by the following set of independent parameters:
\begin{equation}
	M_1, \; M_2, \; \theta_{hS}, \; \theta_{hX}, \; \theta_{SX}, \; v_S, \; v_X .
\end{equation}

We select a set of representative benchmark points after imposing theoretical constraints from perturbative unitarity, vacuum stability, and perturbativity using \texttt{ScannerS} package~\cite{Muhlleitner:2020wwk}, as well as experimental bounds from LEP, Tevatron, and the LHC via the python package \texttt{HiggsTools}~\cite{Bahl:2022igd}. We further refine our selection of benchmark points based on kinematic considerations in particular the mass splitting $M_2 - 2M_1$, production cross sections, and the requirement that $h_1$ predominantly decays into $b\bar{b}$. Three exemplary benchmark points are summarized in Table~\ref{tab:Bps_trsm}. 

\begin{table}[H]
	\centering
	\begin{tabular}{c|ccccccc}
		& $M_{1}$ & $M_{2}$ & $M_{2}-2M_{1}$ & $\sigma(Z h_2)$ & $\sigma(W^\pm h_2)$ & ${\rm Br}(h_2\to h_1 h_1)$ &${\rm Br}(h_1\to b \bar{b})$ \\
		\hline
		BP1 & 20.13 & 42.96 & 2.70  & 0.63 & 1.27 & 0.87 & 0.856\\
		BP2 & 24.84 & 55.25 & 5.56   & 0.37 & 0.71 & 0.93 & 0.866\\
		BP3 & 21.60 & 58.31 & 15.12  & 0.33 & 0.63 & 0.95 & 0.859\\
	\end{tabular}
	\caption{Benchmark points (BPs). Masses are given in \GeV and cross sections in \pb. }
	\label{tab:Bps_trsm}
\end{table}

\section{Collider Analysis}

The target signal in this work is the associated production of a scalar boson with an electroweak gauge boson, $Vh_2$, leading to final states with four $b$-jets accompanied by either missing transverse energy and a charged lepton or by a pair of opposite-sign same flavour leptons.

For the single-lepton channel ($W^\pm  h_2$), the most important background processes are top-quark pair production ($t\bar{t}$), single-top production in the $tW$ channel, and $W$+jets events, including heavy-flavor contributions such as $Wb\bar{b}$. 
For the dilepton channel ($Zh_2$), the main backgrounds arise from $t\bar{t}$ and $Z$+jets production with heavy-flavor components, in particular $Zb\bar{b}$. 
Diboson processes ($WW$, $WZ$, $ZZ$) and $t\bar{t}V$ ($V=W,Z$) constitute additional backgrounds for both channels. 

Signal and background events are generated at leading order using \texttt{MadGraph5\_aMC@NLO}~\cite{Alwall:2014hca} at a center-of-mass energy of 13.6~TeV.  Parton showering and hadronization are performed with \texttt{Pythia~8.3}~\cite{Sjostrand:2014zea}, while detector effects are simulated using \texttt{Delphes~3.5.0}~\cite{deFavereau:2013fsa}.  The DeepCSV algorithm~\cite{CMS:2017wtu} is used as a reference for $b$-tagging performance. In our analysis, we implement only its reported efficiency and mis-tagging rates as functions of transverse momentum, assuming a tagging efficiency of $70\%$ for $b$-jets, a mis-tag rate of $10\%$ for $c$-jets, and $1\%$ for light-flavor jets ($u,d,s,g$).

We apply the following preselection criteria: jets are required to satisfy $p_T > 20$~GeV, while leptons must have $p_T > 10$~GeV. 
All reconstructed objects are restricted to the pseudorapidity range $|\eta| < 2.5$ and must be separated by $\Delta R(i,j) > 0.4$ $(i,j = b,j,\ell)$. 
The production cross sections for signal and background processes at generation level are reported in Tables~\ref{tab:combined} and~\ref{tab:bgxs}, respectively\footnote{For the final results we might consider including higher-order corrections at least for the SM like backgrounds, as these can prove to be quite significant.}.

\renewcommand{\arraystretch}{1.1}
\setlength{\tabcolsep}{0.06cm}
\begin{table}[H]
	\centering
	\begin{tabular}{|c|ccc|cccc|}
		\hline
		\multirow{2}{*}{BP} & \multicolumn{3}{c|}{Masses [\GeV]} & \multicolumn{4}{c|}{Cross-sections $\sigma$ [\pb]} \\
		\cline{2-8}
		& $M_1$ & $M_2$ & $M_2-2M_1$ & $W^+ h_2$ & $W^- h_2$ & $qq(Z h_2)$ & $gg(Z h_2)$  \\
		\hline
		BP1  & 20.13 & 42.96 & 2.70  & 0.10002 (7)  & 0.07025 (5)  & 0.02552 (2)  & 0.006252 (8) \\
		BP2  & 24.84 & 55.25 & 5.56  & 0.06292 (5)  & 0.04360 (3)  & 0.01625 (1)  & 0.005733 (6) \\
		BP3  & 21.60 & 58.31 & 15.12 & 0.05656 (4)  & 0.03907 (3)  & 0.01465 (1)  & 0.005706 (5) \\
		\hline
	\end{tabular}
	\caption{Leading-order cross-sections at the LHC with $\sqrt{s}=13.6$ \TeV, without cuts, using PDF set \texttt{NNPDF23\_nlo\_as\_0119} \cite{Ball:2012cx}. The signals correspond to $4b + \ell^+\nu_\ell$ ($W^+h_2$), $4b + \ell^-\bar{\nu}_\ell$ ($W^-h_2$), and $4b + \ell^+\ell^-$ ($Zh_2$: $q\bar{q^\prime}$ and $gg\to Zh_2$).}
	\label{tab:combined}
\end{table}

\begin{table}[H]
	\centering
	\begin{tabular}{|cccc|ccc|}
		\hline
		\multicolumn{4}{|c|}{Single-Leptons} & \multicolumn{3}{c|}{Di-leptons } \\\hline
		$t\bar{t}$&$tWb$&$W+\text{jets}$&$Wbb$&$t\bar{t}$&$Zbb$&$Zbbj$\\\hline
		$76.48$&$0.02417$&$3221$&$45.39$&$25.46$&$60.94$&$29.66$\\\hline
	\end{tabular}
	\caption{Leading-order background cross sections (in pb) for both production channels at generation level.
	}
	\label{tab:bgxs}
\end{table}

\begin{figure}[htb!]
	\centering
	\includegraphics[width=0.45\linewidth]{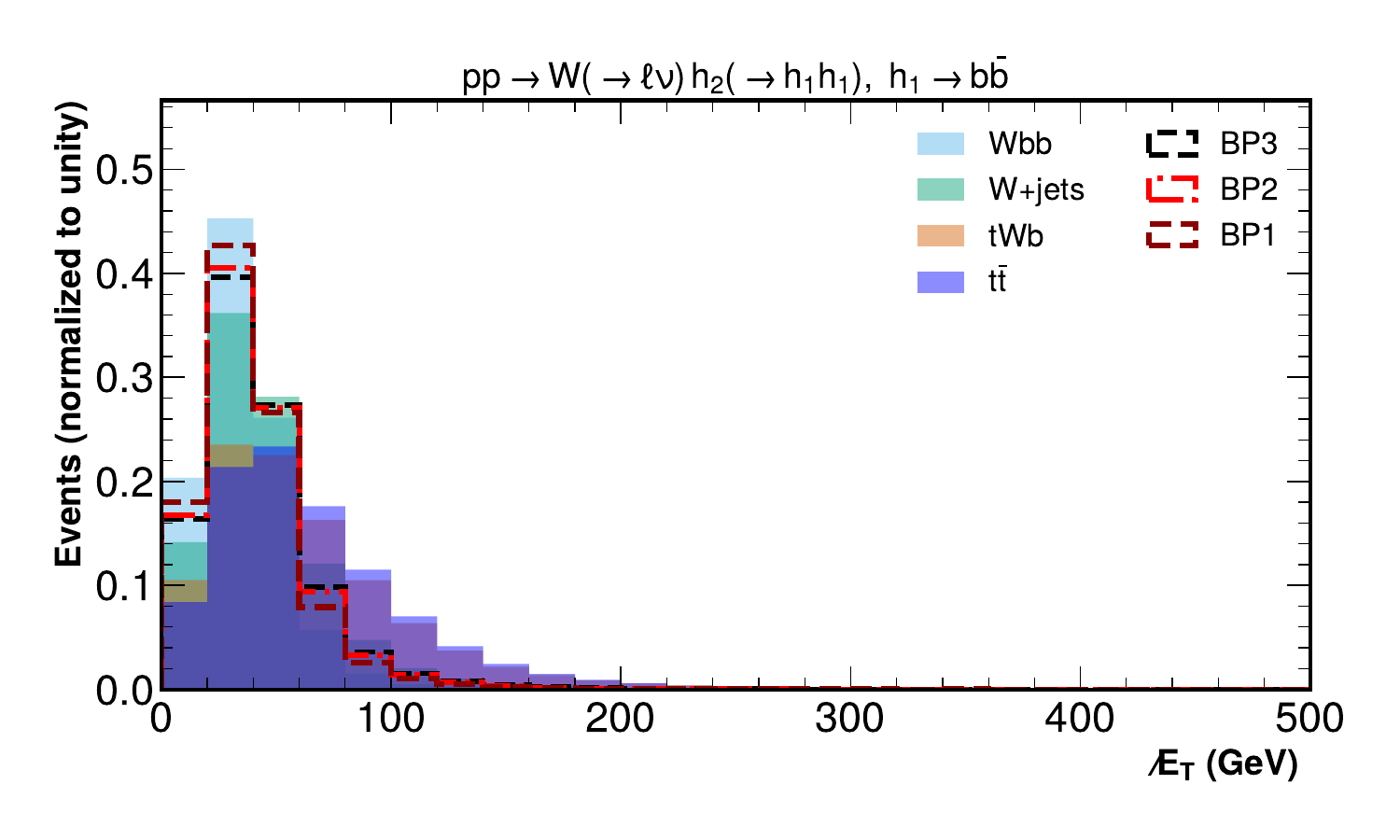}
	\includegraphics[width=0.45\linewidth]{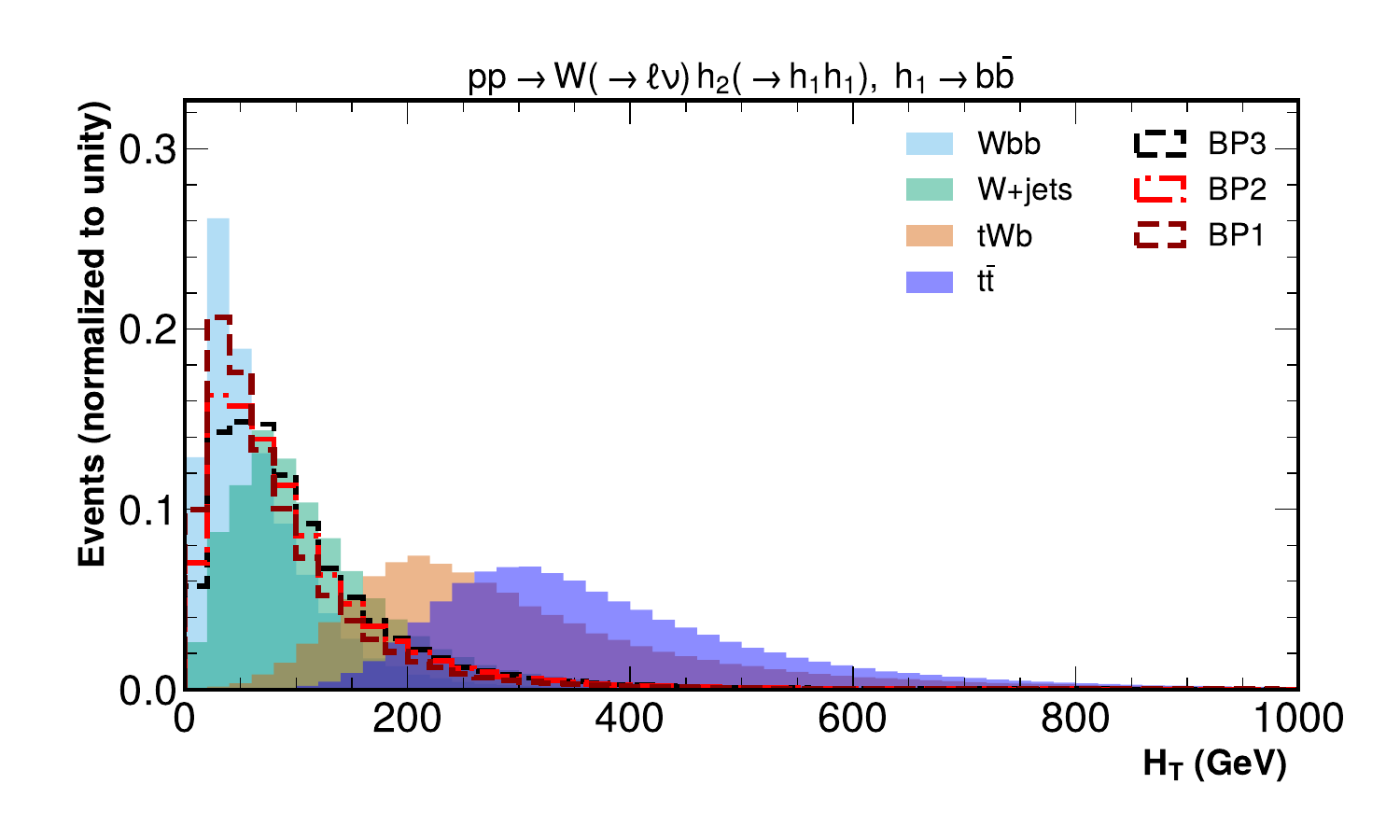}
	\caption{Missing transverse energy (left) and total hadronic transverse energy (right). All distributions are normalized to unity for shape comparison.}
	\label{figs:energy}
\end{figure}

Figure~\ref{figs:energy} shows the distributions of the missing transverse energy and the total hadronic transverse energy for the signal (BPs) and some selected background processes. 
The signal events are predominantly concentrated in the low missing transverse energy region, with $\slashed{E}_T < 100$~GeV, while most background events populate higher values. 
Similarly, the total hadronic transverse energy distribution indicates that the signal is favored in the region $H_T < 200$~GeV, where a better signal-to-background separation is achieved.

\begin{figure}[htb!]
	\centering
	\includegraphics[width=0.45\linewidth]{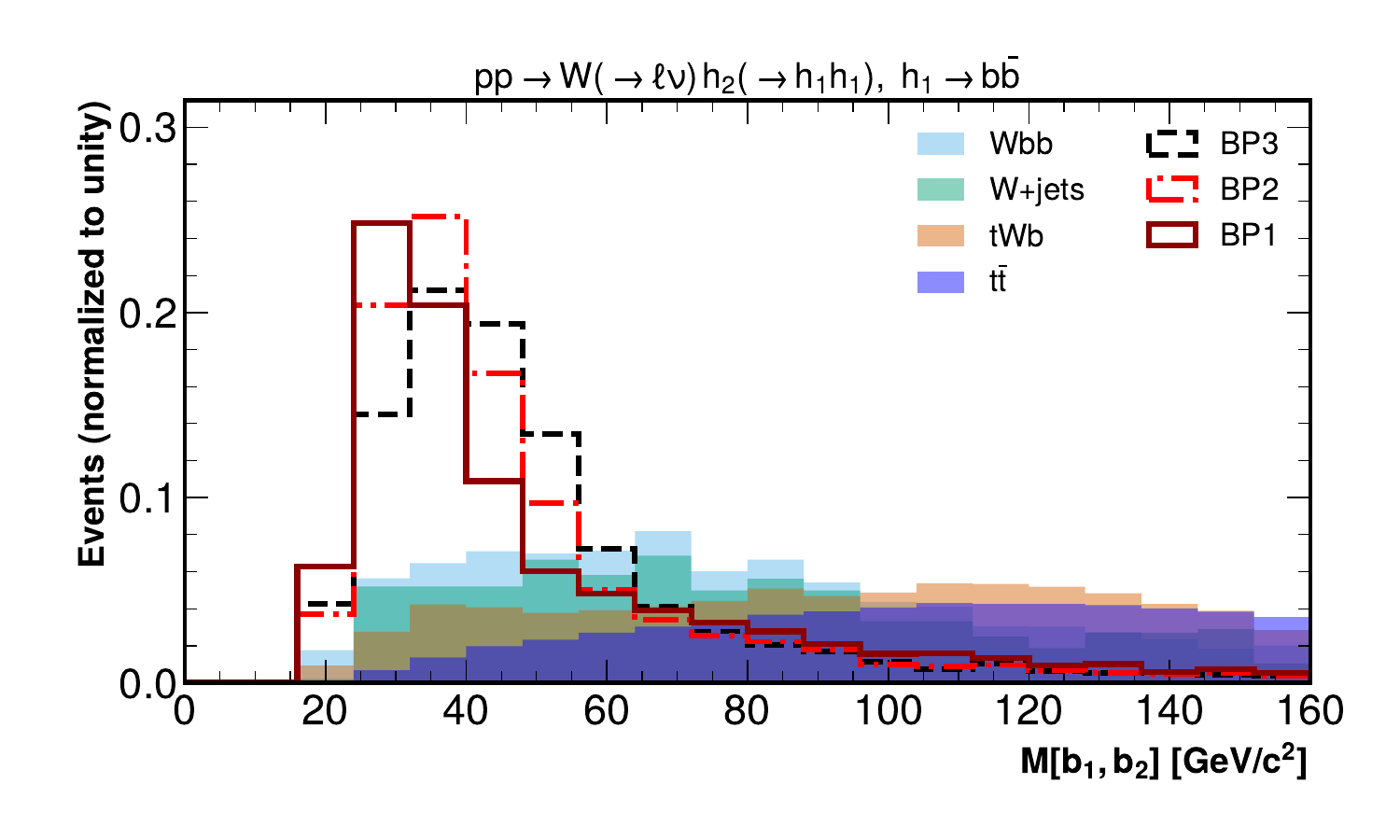}
	\includegraphics[width=0.45\linewidth]{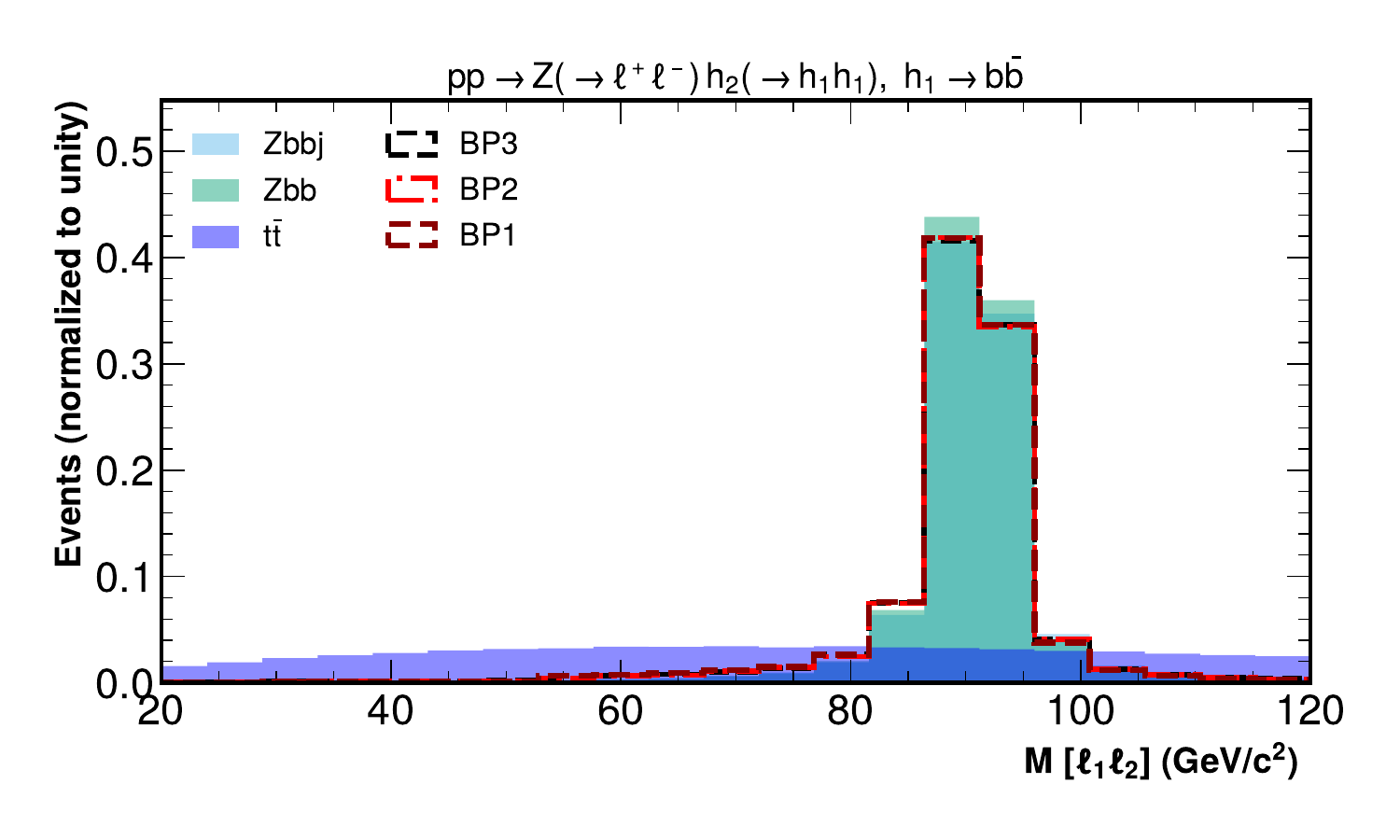}
	\caption{Invariant mass distributions for leading and subleading $b$-jets (left) and leading and subleading leptons (right). All distributions are normalized to unity for shape comparison.}
	\label{figs:Dis}
\end{figure}

Figure~\ref{figs:Dis} shows the invariant mass distributions of the leading and subleading $b$-tagged jets in the $W h_2$ channel (left) and the reconstructed dilepton invariant mass in the $Z h_2$ channel (right). 
These distributions illustrate the reconstruction of the light scalar state $h_1$ and $Z$ boson.

\subsection{Event Selection}

In the $W h_2$ channel, events are required to contain exactly one isolated charged lepton and at least four reconstructed jets, among which at least two must be identified as $b$-tagged jets. 
In the $Z h_2$ channel, events are required to contain at least four reconstructed jets, with at least two of them $b$-tagged. In addition, this channel must include a pair of isolated, oppositely charged leptons.

In both channels, we impose upper bounds on missing transverse energy and total hadronic activity,
\begin{equation}
	\slashed{E}_T (H_T) < 100~(200)~\mathrm{GeV}.
\end{equation}

Given the low-mass and low-$p_T$ regime of the signal, additional kinematic requirements are applied:
\begin{align}
	&p_T(b_1) < 60~\mathrm{GeV}, \quad p_T(b_2) < 40~\mathrm{GeV}, \nonumber\\
	&p_T(j_1) < 60~\mathrm{GeV}, \quad p_T(j_2) < 40~\mathrm{GeV}, \nonumber\\
	&\Delta R (b,\ b) < 1.2, \quad \Delta R (j,\ j) < 1.2,
\end{align}
where all jets are $p_T$ ordered. 

We further require the reconstruction of the light scalar resonance,
\begin{equation}
	|M_1 - M(b,b)| < 10~\mathrm{GeV},
\end{equation}
and impose a mass window around the $Z$ boson mass,
\begin{equation}
	m_Z \in [80,100]~\mathrm{GeV}.
\end{equation}
These selection criteria are optimized to maximize the signal significance.

\subsection{Statistical Analysis}

The discovery significance is evaluated using the profile likelihood approximation~\cite{Cowan:2010js}:
\begin{equation}
	\label{eq:syst}
	\text{SiG}(S,B)=\sqrt{
		2\Biggl[
		(S+B)\ln\Biggl(\frac{(S+B)(B + \sigma^2_B)}{B^2+ (S + B)\sigma^2_B}\Biggr)
		-\frac{B^2}{\sigma^2_B}\ln\Biggl(1 + \frac{\sigma^2_B S}{B(B +\sigma^2_B)}\Biggr)
		\Biggr]
	}.
\end{equation}

Here, $S$ and $B$ denote the expected number of signal and background events, respectively, while $\alpha = \sigma_B/B$ represents the relative systematic uncertainty on the background normalization.

In this analysis, we assume perfect background determination, corresponding to $\alpha=0$. 
The preliminary significances for benchmark points BP1, BP2, and BP3 at an integrated luminosity of $300$ and $3000~\mathrm{fb}^{-1}$ are presented in Table~\ref{tab:sig} for the currently considered backgrounds which are partially statistically limited. 
A more complete study is in progress.

\begin{table}[H]
    \centering
    \begin{tabular}{@{}cccccc@{}}
        \toprule
        \multirow{2}{*}{$\mathcal{L}$ [fb$^{-1}$]} & \multirow{2}{*}{BPs} & \multicolumn{3}{c}{Discovery Significance} \\
        \cmidrule{3-5}
        & & $W^+h_2$ & $W^-h_2$ & $Zh_2$ \\
        \midrule
        \multirow{3}{*}{300} & BP1 & 5.07 & 4.22 & 1.05 \\
        & BP2 & 7.05 & 4.84 & 1.17 \\
        & BP3 & 7.42 & 3.45 & 0.86 \\
        \midrule
        \multirow{3}{*}{3000} & BP1 & 16.03 & 13.33 & 3.32 \\
        & BP2 & 22.30 & 15.31 & 3.69 \\
        & BP3 & 23.47 & 10.92 & 2.72 \\
        \bottomrule
    \end{tabular}
    \caption{Tentative discovery significance for single-lepton and dilepton channels at 13.6~TeV and $300~\mathrm{fb}^{-1}$, and its projection to HL-LHC, for BP1, BP2, and BP3. More detailed studies are in progress to update these numbers.
    }
    \label{tab:sig}
\end{table}

\section{Conclusion}

In this preliminary work, we have investigated the prospects for discovering a light scalar boson in the framework of the Two Real Singlet Model through associated $V h_2$ production at the LHC. We focus on final states consisting of four $b$-jets and leptons. 
We carry out the analysis at $\sqrt{s}=13.6$~TeV. 
After applying selection criteria, we find that several benchmark scenarios might reach discovery-level significance with $300~\mathrm{fb}^{-1}$ of integrated luminosity. 
Our results demonstrate that the $V h_2$ channel provides a complementary and sensitive probe of extended scalar sectors.

\section*{Acknowledgments}
TR and MO are supported by the Croatian Science Foundation (HRZZ) under Grant No. HRZZ-IP-2022-10-2520. The authors acknowledge support from the COMETA COST Action CA22130. PS is supported by Basic Science Research Program through the National Research Foundation of Korea (NRF) funded by the Ministry of Education through the Center for Quantum Spacetime (CQUeST) of Sogang University with grant number RS-2020-NR049598 and by the Ministry of Science and ICT with grant number RS-2025-24523022.

\bibliographystyle{JHEP}
\bibliography{ref}
\end{document}